\documentclass[times,twocolumn,final,authoryear]{elsarticle}

\usepackage{jasr}
\usepackage{framed,multirow}

\usepackage{amssymb}
\usepackage{latexsym}
\usepackage{amsmath}
\usepackage{dsfont}

\usepackage[switch]{lineno}

\usepackage{url}
\usepackage{xcolor}
\definecolor{newcolor}{rgb}{.8,.349,.1}

\usepackage[citebordercolor=white]{hyperref}

\journal{Advances in Space Research}

\begin{document}

\verso{Heyns \textit{et al.}}

\begin{frontmatter}

\title{Adaptations to a geomagnetic field interpolation method in Southern Africa}

\author[1,2]{M. J. \snm{Heyns}\corref{cor1}}
\cortext[cor1]{Corresponding author}
\ead{m.heyns@imperial.ac.uk}
\author[3]{S. I. \snm{Lotz}}
\author[3]{P. J. \snm{Cilliers}}

\author[2]{C. T. \snm{Gaunt}}

\address[1]{Department of Physics, Imperial College London, London, UK, SW7 2BW}
\address[2]{Department of Electrical Engineering, University of Cape Town, Cape Town, South Africa, 7700}
\address[3]{South African National Space Agency, Hermanus, South Africa, 7200}

\received{17 November 2021}
\finalform{8 March 2022}
\accepted{xx March 2022}
\availableonline{xx March 2022}
\communicated{}

\begin{abstract}
Space weather and its impact on infrastructure presents a clear risk in the modern era, as evidenced by the adverse effects of geomagnetically induced currents (GICs) in power networks. To model GICs, ground-based geomagnetic field (B-field) measurements are critical and need to be available in the region of interest. A challenge globally lies in the sparse distribution of magnetometer arrays, which are seldom located near critical power network nodes. Interpolation of the geomagnetic field (B-field) is often needed, with the spherical elementary current system (SECS) approach developed for high-latitude regions favoured. We adapt this interpolation scheme to include low-cost variometers to interpolate $dB/dt$ directly and increase interpolation accuracy. A further adaptation to the scheme is to physically represent the mid-latitude context where most power networks and pipelines lie. The driving current systems in these regions differ from their high-latitude counterparts. Using a physics-consistent mid-latitude version of SECS, we show why previous implementations in Southern Africa are incorrect but still result in useful interpolation. The scope of these adaptations not only has direct application to research in general, but also to utilities, where effective low-cost instrumentation can be used to improve GIC modelling accuracy.
\end{abstract}


\end{frontmatter}

%% For linenumbers
%\linenumbers

\section{Introduction}
\label{sec:intro}
In the context of increasing societal dependence on technology, the need for improved space weather awareness has been recognised internationally. Loosely, space weather is the effect of solar activity on the Earth and the near-Earth environment. The technologies affected by adverse space weather are widespread, ranging from satellite operation, HF communication and GPS navigation, to grounded conducting networks such as power grids, railways and pipelines \citep{Boteler1998,Kelbert2020,Boteler2020}. The primary space weather related cause of damage in power grids and pipelines arises from geomagnetically induced currents (GICs). In power networks, GICs driven by extreme space weather events may result in total system collapse, similar to that which led to the 1989 Hydro-Qu\'{e}bec blackout \citep{Bolduc2002,Boteler2019}. A conservative estimate of the direct economic impact of this event, which caused a 12-hour long power blackout for millions of people, was on the order of \$13.2 million \citep{Bolduc2002}. The indirect and knock-on economic effects may be much larger \citep{Oughton2017}. In response to the threat posed by GICs, the United States Federal Energy Regulatory Commission (FERC) issued a directive for all utilities in the United States to assess GIC risks and undertake data collection in support of research in the field \citep{FERC}. GIC research efforts however have been international, and include engineering aspects \citep{Divett2018}, geophysical modelling \citep{Love2016,Marshall2020} and economic assessments \citep{Oughton2017,Akpeji}.

For most space weather ground-effects, it is the interaction of solar plasma with the near-Earth environment that acts as the driving mechanism. Short-term fluctuations of the geomagnetic field (B-field) measured on the ground are a direct result of these interactions. The Earth’s roughly dipolar B-field is perturbed by near-Earth currents, created by interactions between the solar wind plasma and the magnetosphere. As such, geomagnetic activity as measured by the B-field on the surface of the Earth generally tracks space weather activity. For GICs, the B-field and its fluctuations ($dB/dt$) are more than just a proxy for space weather activity. Through Faraday’s law of induction, the fluctuations of the penetrating B-field induce a geoelectric field (E-field) and telluric currents within Earth. These naturally occurring currents are often used by the geophysics community to estimate the subsurface conductivity structure through magnetotelluric (MT) modelling in the frequency domain. When a grounded conducting network is present, the induced currents may enter the network as GICs and disrupt operations. Accurate estimation of the disturbance B-field or $dB/dt$ is the first step in providing accurate modelling and forecasting of GICs.

A challenge globally, for both GIC and other geophysical research, is the sparse spatial coverage of observatory-grade magnetometers that can measure the absolute values of the geomagnetic field. A natural approach has been to make use of spatial interpolation of the B-field. Various interpolation methods have been used, ranging from simplistic nearest neighbour approaches to spherical cap harmonic analysis \citep{Torta2020}, Delauney triangulation \citep{Torta2017} and polynomial interpolation using magnetic scalar potentials \citep{Duzgit1997}. The favoured method is however the physically inspired spherical elementary current systems (SECS) interpolation scheme \citep{Amm1997,Amm1999}. The basic idea behind SECS interpolation is that an equivalent current system at a height similar to a driving near-Earth current system can be estimated using basis functions, with coefficients constrained using B-field measurements at geomagnetic observatories. At high-latitudes, where the complex auroral electrojets occur, the assumed height is around 100 km, which locates the equivalent currents in the D-region of the ionosphere. Further, it is assumed that at high-latitudes divergence-free current nodes can be used as basis functions to fully describe the ionospheric Hall currents associated with near-vertical field-aligned currents. In previous GIC modelling in Southern Africa, SECS has been used extensively to interpolate B-field estimates within the power network and improve on GIC estimates \citep{Bernhardi2008,Ngwira2009,Matandirotya2015}. However, for the mid-latitude Southern African context where distant magnetospheric currents are the main driver, divergence-free basis functions are not expected to be representative of the more spatially consistent B-field. Explicitly, the use of divergence-free basis functions has been identified as a cause of inadequate interpolation accuracy using SECS in other mid-latitude regions \citep{McLay2010,Torta2017}. 

This paper acts as a follow-up to the original implementation of SECS in Southern Africa by \citet{Bernhardi2008} and will address two important adaptions in B-field interpolation that improve interpolation accuracy and interpretability. The first is to include data from low-cost fluxgate magnetometers operated as variometers, i.e. not observatory-grade magnetometers, to improve interpolation accuracy. The second adaption is to correct the previous implementation of SECS and then identify basis functions that are more appropriate for mid-latitudes. 

The next section outlines the extent of magnetometer measurements in Southern Africa, with datasets identified for validation of SECS interpolation and its adaptations. The mechanics of SECS interpolation and the proposed adaptations are then presented, followed by implementation of both in the results section. The paper concludes with a discussion of the results and the implications of adapted B-field interpolation.

\section{Geomagnetic Field Measurements in Southern Africa}
\label{sec:meas}

\begin{figure}
\centering
\includegraphics[scale=0.125]{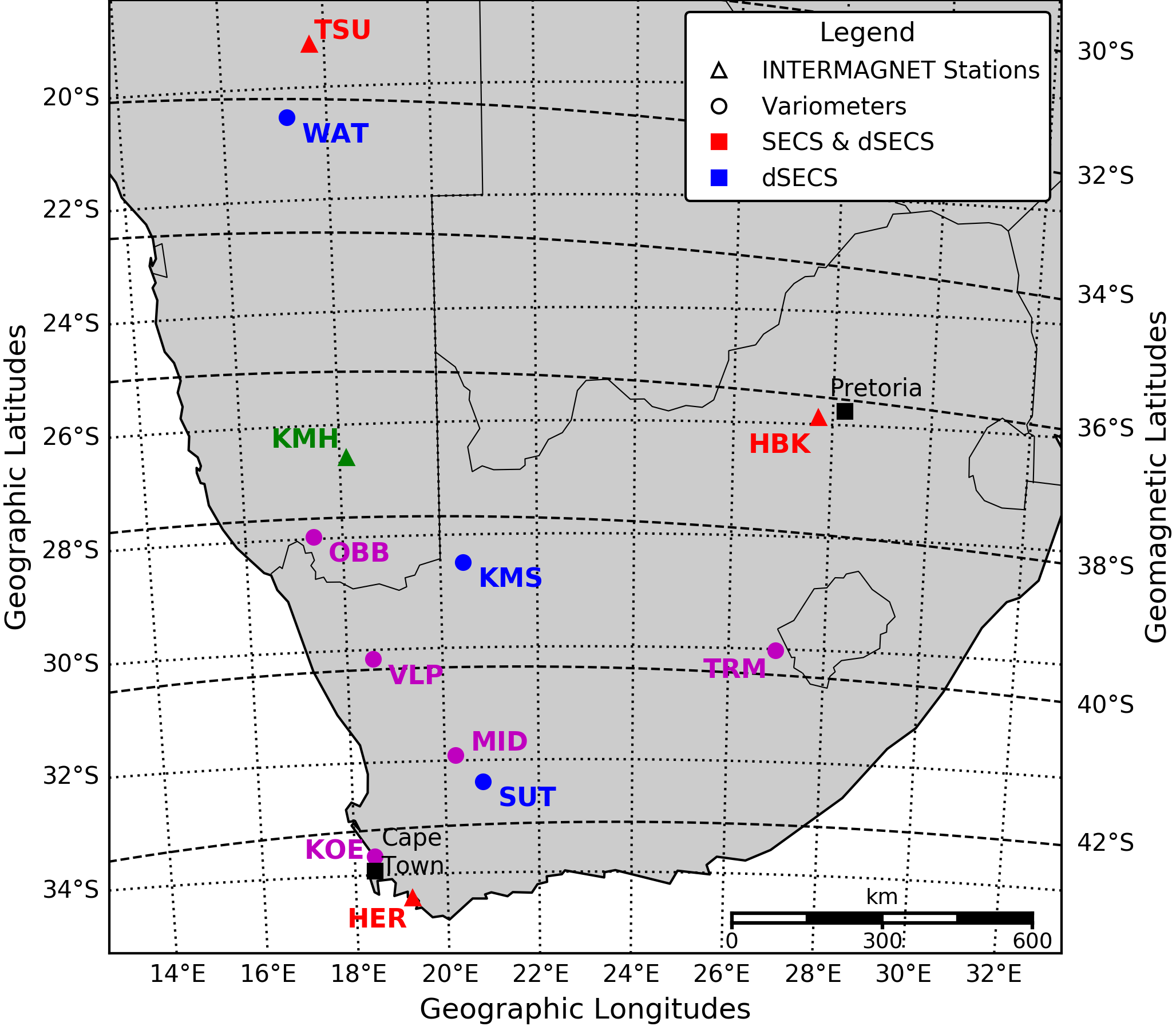}
\caption{Geomagnetic measurements in Southern Africa. There are four INTERMAGNET sites (HER, HBK, KMH, TSU), two magnetometer sites (WAT, SUT) and eight magnetotelluric station magnetometers (KMS, OBB, VLP, MID, KOE, TRM, HER, HBK). Data from KMH (green) are used for validation purposes in this study. Data from the INTERMAGNET sites (red) drive conventional SECS interpolation and data from variometer sites (blue) are additionally included to drive the adapted dSECS interpolation method. Geomagnetic coordinates are estimated using a quasi-dipole approximation at epoch 2015.0 \citep{Emmert2010}. Locations are further summarised in Table \ref{tab:sites}.}
\label{fig:map}
\end{figure}

Southern Africa has a long legacy of geomagnetic research and currently has four geomagnetic observatories (HER, HBK, KMH and TSU), operated by the South African National Space Agency as part of the international INTERMAGNET network (\url{www.intermagnet.org}). Although these geomagnetic observatories form the backbone of geomagnetic research, there is an increasing number of field assets with similar potential. Specifically, additional B-field measurements may come from lower cost magnetometers that are not observatory-grade (often not temperature controlled, which may introduce drift that affects measurements when not corrected for). Similar to variometers or induction magnetometers, these low-cost magnetometers are able to accurately measure B-field variation (per timestep). As shown in Figure \ref{fig:map}, leveraging these additional measurements would create a much denser grid of B-field measurements. 

Magnetometer data from the various sites in Figure \ref{fig:map} and Table \ref{tab:sites} are used for  both interpolation and validation. First, there is initial validation of the SECS interpolation scheme as applied by previous work. Validation was done using the INTERMAGNET sites to drive interpolation across Southern Africa, which is then compared to B-field measurements from sites where LEMI-417M magnetotelluric (MT) sensors are deployed. The second use of data is in the validation of adaptations of the interpolation scheme. These adaptations require concurrent data coverage across sites and, due to availability, four geomagnetic storms in 2015 are used (see Table \ref{tab:storms}, 22,224 data points in total). The data used include samples from all phases of the geomagnetic storm, from quiet time to the impulsive peak, rapidly varying main phase and slowly varying recovery phase. High-fidelity B-field measurements from the KMH observatory are kept as the out-of-sample dataset for validation purposes. Adaptations are driven with data from either the remaining three INTERMAGNET sites (HER, HBK, TSU), or additionally including data from magnetometers at SUT, WAT and KMS. The observatories at SUT and WAT make use of LEMI-025 3-axis fluxgate magnetometers, whereas KMS is an MT site equipped with a LEMI-417M 3-axis fluxgate magnetometers and a 2-axis E-field sensor. All fluxgate instruments are designed to measure the absolute magnetic field components along three orthogonal axes. The absolute field measurements are however subject to drift, which is corrected through manual baseline measurements (as done at INTERMAGNET sites). In this paper, for ease of reference we will refer to low-cost fluxgate magnetometers with uncorrected baselines as variometers. More accurately, a variometer is used to measure the variation of the field components about baseline values, usually through induction measurements of $dB/dt$. Induction magnetometers are typical examples of variometers. These magnetometers are based on Faraday's law of induction and use induction coils to measure the emf (voltage) induced by changes in magnetic flux through the coil, i.e. $dB/dt$ and not the absolute B-field. 

\begin{table}
\centering
\caption{Geomagnetic measurement sites in Southern Africa (see Figure \ref{fig:map}).}\label{tab:sites}
\begin{tabular}{c|c|cc|cc}
Station &Station & \multicolumn{2}{c}{Geographic}& \multicolumn{2}{|c}{Geomagnetic}\\
\underline{Type} / Name & Code &  Lat. &   Lon. &   Lat. &   Lon. \\
\hline
\hline
\underline{INTERMAGNET} & &  &  &  & \\
Hartebeesthoek & HBK & -25.9 & 27.7 & -36.3 & 97.0 \\
Hermanus & HER & -34.4 & 19.2 & -42.8 & 84.6 \\
Keetmanshoop & KMH & -26.6 & 18.1 & -37.2 & 87.0 \\
Tsumeb & TSU & -19.2 & 17.7 & -31.1 & 89.2 \\
\underline{LEMI-025} & &  &  &  & \\
Sutherland & SUT & -32.4 & 20.8 & -41.5 & 87.2 \\
Waterberg & WAT & -20.5 & 17.2 & -32.3 & 88.3 \\
\underline{LEMI-417M} & &  &  &  & \\
Kakamas & KMS & -28.5 & 20.5 & -38.6 & 88.7 \\
Koeberg & KOE & -33.4 & 18.3 & -42.3 & 84.2 \\
Middelpos & MID & -31.5 & 20.1 & -41.1 & 86.8 \\
Obib & OBB & -27.5 & 16.4 & -38.2 & 85.7 \\
Trompsberg & TRM & -30.0 & 26.0 & -39.7 & 94.5 \\
Vaalputs & VLP & -30.1 & 18.3 & -39.9 & 85.9 \\
\end{tabular}
\end{table}

\begin{table}
\centering
\caption{Data availability for HER, HBK, TSU, SUT, WAT and KMS sites to validate interpolation adaptations relative to KMH. Maximum Kp (planetary K-index) during periods is included as an indication of geomagnetic activity.}\label{tab:storms}
\begin{tabular}{c|c|c}
  Geomagnetic &   Data &   Max\\
  Storm &   Range &   Kp \\
\hline
\hline
1 & 15-16, 18 March 2015 & 6 \\
%\hline
2 & 21-23 June 2015 & 8+ \\
%\hline
3 & 09-12 September 2015 & 7 \\
%\hline
4 & 06-14 September 2015 & 7+ \\
\multicolumn{3}{c}{}\\
\hline
\end{tabular}
\end{table}

\section{SECS Interpolation}
\label{sec:inter}

\begin{figure}
\centering
\includegraphics[scale=0.35]{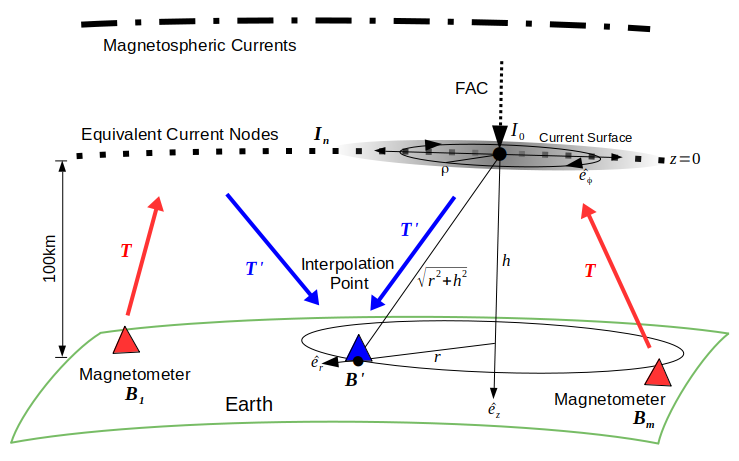}
\caption{To represent Hall currents associated with an vertical field-aligned current (FAC), the divergence-free SECS interpolation scheme uses an equivalent current node. This current node has an amplitude $I_0$, which sets up a surface current density defined by $\vec{J_{df}}(\rho)$ on a current surface 100 km above the Earth's surface. The Biot-Savart mapping $T$ allows magnetometer measurements $B_m$ to constrain a current system of multiple current nodes $I_n$, which then can be used to interpolate the horizontal B-field at $B'$ through the inverse mapping $T'$.}
\label{fig:secs}
\end{figure}

SECS uses an equivalent current system to create a physically inspired interpolation scheme based on Maxwell's equations. As shown by \citet{Amm1999}, including this prior adds significant robustness when compared with other purely mathematical interpolation schemes that do not consider the physical context. From Helmholtz's theorem, any current flowing on a surface can be broken into a curl-free part (allowing current flow into and out of the surface) and a divergence-free part (allowing current flow on the surface). In terms of the real-world current systems at play, overhead ionospheric Hall currents associated with the auroral electrojet system would be linked to the divergence-free part. The curl-free part is more often associated with field-aligned and Pedersen currents. Assuming a uniform conductivity of the ionosphere and field-aligned currents that are perpendicular to the ground, ground-based measurements would predominantly measure the overhead divergence-free Hall currents at high-latitudes \citep{Fukushima1976}. SECS interpolation aims to approximate a similar real-world current system with an equivalent current system or current surface at an arbitrary height, as shown in Figure \ref{fig:secs}. 

Of the different SECS implementations, we assume cylindrical coordinates ($r=\sqrt{x^2+y^2},\phi,z$) \citep{Viljanen2004}, with a current element of amplitude $I_0$ at height $h$, as shown in Figure \ref{fig:secs}. Implementation of SECS at mid-latitudes assumes that divergence-free current system representation still holds, and an equivalent current height of 100 km has been used in previous implementations in Southern Africa \citep{Bernhardi2008,Ngwira2009}. It should be noted that no physical current system exists at this height for mid-latitudes. Rather the dominant current systems are the more distant magnetospheric currents \citep{DeVilliers2017}. Nevertheless, for divergence-free current nodes $I_0$ within the SECS interpolation scheme, the equivalent surface current density $\vec{J_{df}}(\rho)$ as a function of radial distance $\rho$ from current node on the current surface is,

\begin{eqnarray}
\vec{J_{df}}(\rho)=\frac{I_0}{2\pi\rho}\hat{e_\phi},
\end{eqnarray}
where $\hat{e_\phi}$ is in the azimuthal direction. 

Choosing a cylindrical coordinate system where $z$ is downwards, and assuming the disturbance B-field shows harmonic time dependence (i.e. $e^{i\omega t}$), a current node $I_0$ can then be related to a measured B-field on the ground for a given time instance through \citep{Viljanen2004},

\begin{eqnarray}
\vec{B}(r)=\frac{\mu_0I_0}{4\pi r}\left(\left(1-\frac{h}{\sqrt{r^2+h^2}}\right)\hat{e_r}+\left(\frac{r}{\sqrt{r^2+h^2}}\right)\hat{e_z}\right).\label{eq:secs}
\end{eqnarray}

Here $\mu_0$ is the permeability of free-space, $r$ is the radial distance from the current node to observer on the surface of the Earth and $h$ is the height of the equivalent current system.

For some defined spatial extent, we can then extend this Biot-Savart relation from a single current element to a grid of elementary current elements. Taking the current grid (say $n$ elements), we group all current elements into a vector $I$. To constrain the elementary currents, we require B-field measurements. Assuming there are multiple magnetometers (say $m$ stations), we can similarly group the B-field measurements into a vector $B$. To solve for $I$, a matrix equation is set up, i.e. $B=T\cdot I$, where $T$ is a transfer function matrix that relates the elementary currents to the measured B-field, encoding the geometry of the problem. Since we are only interested in the horizontal B-field, only the radial component $\hat{e_r}$ of \eqref{eq:secs} is used. An important property of the transfer function is that it is purely a spatial relationship between the elementary current and the magnetometer station. 

Diverging from the traditional SECS formalism, but following previous implementations in Southern Africa, the matrix equation calculation is performed separately for the $x$ (north-south) and $y$ (east-west) components of $I$ and $B$ respectively. More explicitly in matrix form,

\begin{eqnarray}
\begin{array} {rcl}
\left[ \begin{array}{c} 
B_{x,y:1} \\ 
\vdots \\ 
B_{x,y:m} 
\end{array} \right] 
& = & 
\left[ \begin{array}{ccc} 
T_{11} & \cdots & T_{1n} \\
\vdots & \ddots & \vdots \\
T_{m1} & \cdots & T_{mn}
\end{array} \right]
\left[ \begin{array}{c} 
I_{x,y:1} \\ 
\vdots \\ 
I_{x,y:n} \end{array} 
\right],
\end{array}
\end{eqnarray}
where,
\begin{eqnarray}
T_{ij}=\frac{\mu_0}{4 \pi r}\left(1-\frac{h}{\sqrt{r_{ij}^2+h^2}}\right).
\end{eqnarray}

In this particular formalism, the $B_{x,y}$ components are related to corresponding current nodes, i.e. $I_{x,y}$ does not refer to direction. The indices $i$ and $j$ refer to the mapping between magnetometer ($[B_1,...,B_m]$) and current element ($[I_1,...,I_n]$) respectively.

Since interpolation is usually needed for sparse magnetometer grids, $m<n$ and the problem is underdetermined. This further means that $T$ is non-square ($m\neq n$), and $I$ needs to be calculated through a quasi-inverse $T^{-1}$, i.e. $I=T^{-1}B$. Commonly, this quasi-inverse $T^{-1}$ is obtained through singular value decomposition (SVD). After successfully solving the inversion problem and estimating the vector $I$ from the measured B-field, $I$ can be used to interpolate the B-field to any other point of interest, i.e. $B'=T'\cdot I$. This procedure is then repeated for all required time instances.

\begin{table}
\centering
\caption{Comparison of MT variometer measurements (LEMI-417M magnetotelluric sensors) with geomagnetic observatories and SECS interpolation over Southern Africa as defined by RMSE and the correlation coefficient $\rho$.}\label{tab:val}
\begin{tabular}{c|c|c|c|c}
Site & Data Range & Comparison & \it RMSE & $\rho$\\
\hline
\hline
HER & 5/4/2012-15/8/2012 & INTERMAG & 16 nT & 0.82 \\
& 8/1/2015-27/1/2015 & &  & \\\hline
HBK & 1/11/2013-7/1/2015 & INTERMAG & 16 nT & 0.68 \\\hline
KMS & 1/9/2013-19/10/2015 & SECS & 4 nT & 0.99 \\\hline
MID & 6/7/2012-7/11/2012 & SECS & 4 nT & 0.99 \\\hline
OBB & 30/8/2013-11/12/2014 & SECS & 8 nT & 0.93 \\
& 17/9/2015-11/12/2015 & & & \\\hline
VLP & 22/5/2012-9/11/2012 & SECS & 17 nT & 0.76 \\
& 25/5/2014-14/2/2015 & & &
\end{tabular}
\end{table}

\subsection{Validation of Previous SECS Implementation}
\label{subsec:val}

To test the accuracy of the SECS interpolation scheme as implemented in Southern Africa previously, a set of data from MT magnetometers can be used to compare interpolation with measurement. Table \ref{tab:val} below shows various periods of measured B-field at MT sites compared with SECS interpolation at those sites. Of the MT sites used, HER and HBK are co-located with INTERMAGNET stations and the accuracy of MT variometer measurements can be calibrated against observatory-grade magnetometer measurements in general. The interpolation in this case makes use of a current node grid spanning an area larger than the interpolation region, with distances between current nodes roughly 100 km. To mirror previous work done in Southern Africa \citep{Bernhardi2008,Ngwira2009,Matandirotya2015}, the height of the equivalent current system is similarly taken as 100 km and B-field data from HER, HBK, KMH and TSU INTERMAGNET sites are used to drive the interpolation. From the resulting comparisons in Table \ref{tab:val}, the difference in B-field measurements at co-located INTERMAGNET observatory magnetometers and MT magnetometers is larger than the difference between SECS interpolation and similar MT magnetometers at remote sites. The comparison suggests that the previous implementation of the planar SECS interpolation scheme does just as well as having a variometer available - in effect validating previous use of the SECS method for GIC applications in Southern Africa. A question raised by this analysis is why the planar SECS interpolation scheme does so well in Southern Africa, and not at other mid-latitudes where previously implemented \citep{McLay2010,Torta2017}? Possible answers to this question are explored in Sections \ref{sec:corr} and \ref{sec:disc}.

\section{dSECS: Interpolating $dB/dt$ Directly}
\label{sec:dSECS}

If we shift the focus to modelling the variation of the B-field, we do not need the luxury of drift-corrected observatory-grade magnetometers. Instead, we can adapt the principle behind SECS interpolation and use low-cost fluxgate magnetometers or variometers that accurately measure the change in the B-field i.e. $dB/dt=B(t_i)-B(t_{i-1})$. Since the transfer matrix $T$ is time independent, we see that $dB/dt$ can be interpolated in much the same way as $B$. The only difference in this case is that current nodes used for interpolation become current node variations $dI/dt$, i.e.,

\begin{eqnarray}
dB/dt&=&B(t_i)-B(t_{i-1})\nonumber\\
&=&T\cdot I(t_i)-T\cdot I(t_{i-1})\nonumber\\
&=&T\cdot \left(I(t_i)-I(t_{i-1}) \right) \nonumber\\
&=&T\cdot dI/dt.
\end{eqnarray}

Observatory-grade magnetometers can of course act as variometers, with the implication that there will be more variometers available than absolute magnetometers. Given a larger number of variometer measurements, the confidence in $dB/dt$ interpolation would be higher than for the case where only measurements of $B$ are used. The rate of change $dB/dt$ is also what is typically used for geoelectric studies and can be used as an alternative to the B-field in the frequency domain. Using the greater interpolation accuracy of $dB/dt$, we can nevertheless improve the interpolation of $B$. To do this, we consider the time series of $B$ and of $dB/dt$. Let us assume the time series $B$ is of length $N+1$ and hence $dB/dt$ is of length $N$. Assuming there exists an $N+1$ set of perturbations $\epsilon$ that link the interpolated B-field estimate with the more accurate interpolated $dB/dt$ variation estimate, the two resulting time series can be equated,

\begin{eqnarray}
dB_1/dt&=&(B_2+\epsilon_2)-(B_1+\epsilon_1)\nonumber\\
&\cdots&\nonumber\\
dB_N/dt&=&(B_{N+1}+\epsilon_{N+1})-(B_N+\epsilon_N).
\end{eqnarray}

In matrix form $A\vec{x}=\vec{b}$, this would be rewritten as,

\begin{eqnarray}
\hspace*{-25pt}
\small
\textit{\tiny N}\overset{\xrightarrow[\hphantom{\hspace*{3cm}}]{\textit{N+1}}}{\left\downarrow {\footnotesize \left[
\begin{array}{ccccc} 
-1 & 1 & 0 & \cdots & 0 \\
0 & -1 & 1 & \cdots & 0 \\
\vdots & \ddots & \ddots & \ddots & \vdots \\
0 & \cdots & 0 & -1 & 1
\end{array} \right]}
\right.} 
\left[ \begin{array}{c} 
\epsilon_1 \\ 
\epsilon_2 \\ 
\vdots \\ 
\epsilon_{N+1} \end{array} 
\right]=\left[ \begin{array}{c} 
\frac{dB_1}{dt} + B_1 - B_2 \\ 
\frac{dB_2}{dt} + B_2 - B_3 \\
\vdots \\ 
\frac{dB_N}{dt} + B_N - B_{N+1}  
\end{array} \right].
\end{eqnarray}

Although the sparse matrix $A$ is non-square ($N \times N+1$), it has an analytical quasi-inverse that is well-behaved does not need to be estimated through SVD. This inverse is given by,

\begin{eqnarray}
A^{-1}=
\overset{\xrightarrow[\hphantom{\hspace*{5cm}}]{\textit{N}}}{\left.\left[ \begin{array}{ccccc} 
-\frac{N}{N+1} & -\frac{N-1}{N+1} & -\frac{N-2}{N+1} & \cdots & -\frac{1}{N+1} \\
\frac{1}{N+1} & -\frac{N-1}{N+1} & -\frac{N-2}{N+1} & \cdots & -\frac{1}{N+1} \\
\frac{1}{N+1} & \frac{2}{N+1} & -\frac{N-2}{N+1} & \cdots & -\frac{1}{N+1} \\
\vspace{-0.25cm} & & & & \\
\vdots & \ddots & \ddots & \ddots & \vdots \\
\vspace{-0.25cm} & & & & \\ 
\frac{1}{N+1} & \cdots & \frac{N-2}{N+1} & -\frac{2}{N+1} & -\frac{1}{N+1} \\
\frac{1}{N+1} & \cdots & \frac{N-2}{N+1} & \frac{N-1}{N+1} & -\frac{1}{N+1} \\
\frac{1}{N+1} & \cdots & \frac{N-2}{N+1} & \frac{N-1}{N+1} & \frac{N}{N+1}
\end{array} \right]\right\downarrow}\textit{\tiny N+1}\hspace*{2pt}.
\end{eqnarray}

To verify the nature of the quasi-inverse, it can be shown that $AA^{-1}=\mathds{1}(N\times N)$ and $A^{-1}A\approx\mathds{1}(N+1\times N+1)$ when $N\gg1$. Since all the components in $\vec{b}$ are known, this quasi-inverse approach is used again to solve for the perturbations $\vec{x}$. After $\vec{x}$ is estimated in our particular case, each original interpolated baseline $B_i$ time instance is updated by its corresponding perturbation $\epsilon_i$, i.e. the resulting initial B-field interpolation is constrained by the more accurate interpolated $dB/dt$. This formalism of efficiently constraining a baseline estimate by a more accurate variation estimate has widespread application.

\section{Correction to Previous SECS Implementation}
\label{sec:corr}

A possible reason for discrepancy regarding improved performance relative to other SECS implementations in mid-latitude regions lies in the implementation of the interpolation scheme. Specifically, the previous implementations in Southern Africa assumed that the $B_x$ and $B_y$ components can be computed independently with corresponding $I_x$ and $I_y$ current nodes. This is not a correct physical interpretation of the SECS methodology that assumes a single $I_0$ current node related to a horizontal radial B-field ($B_r$), that would need to be projected into its $B_x$ and $B_y$ components. When attempting a correct implementation of SECS using that same data as in the validation shown in Table \ref{tab:val}, the interpolation performance breaks down as the divergence-free current node bases fail to link to the sparse magnetometer measurements. Furthermore, performance is exceptionally sensitive to choice of current system grid as the B-field contributions from various divergence-free current nodes need to cancel to describe a relatively spatially consistent B-field, typical of mid-latitudes.

Although the previous implementation of SECS in Southern Africa diverges from the physical basis of traditional SECS interpolation, the fact that it achieves accurate interpolation suggests a similar adaptation would be feasible. The accuracy in describing a spatially smoother field is due to artificial smoothing of the interpolation across an area much larger than the defined divergence-free equivalent current height of 100 km at mid-latitudes, where additionally no such current system would exist. This smoothing introduced by $(1-h/\sqrt{r^2+h^2})$ is compounded by allowing the two components of the B-field to be defined separately. The combination allows a smooth current system to be represented, when Hall-like ionospheric currents should be the basis. Hence, although not physically consistent, the previous implementation of SECS interpolation in Southern Africa can be seen an effective mathematical transform.

\begin{figure}
\centering
\includegraphics[scale=0.25]{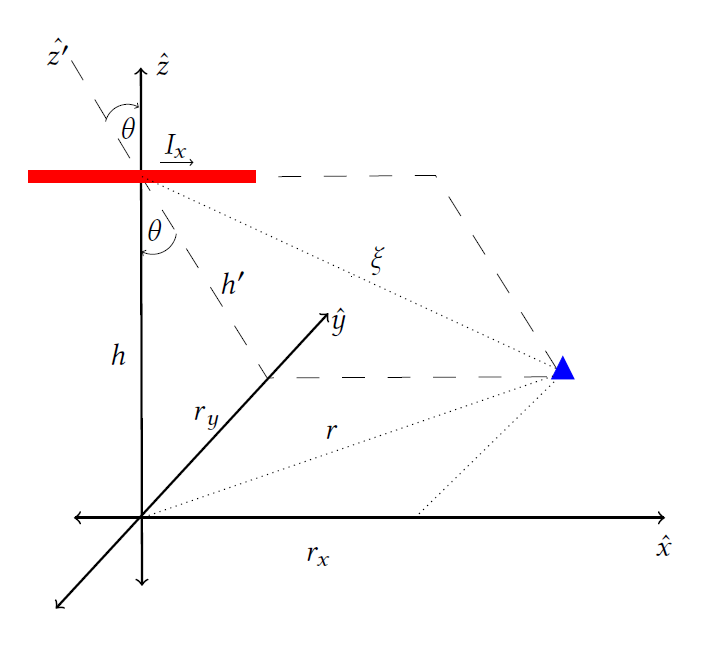}
\caption{Setting up the line element implementation of the equivalent current system (ECS) interpolation, we assume a line current element $I$ at height $h$, a distance $\xi$ away from a ground node. Assuming the current is in the $\hat{x}$ direction, the associated B-field would be in the $\hat{y}$ and/or $\hat{z}$ directions. To change the reference frame, we define the calculation plane ($\theta$ away from the $xz$-plane) that includes both line current element and ground node (dotted lines). In this frame, the current would be at height $h'$ instead and the B-field orthogonal to the calculation plane. A similar orthogonal set-up can be assumed for the $\hat{y}$ direction with associated B-field in the $\hat{x}$ and/or $\hat{z}$ directions.}
\label{fig:corr1}
\end{figure}

\begin{figure}
\centering
\includegraphics[scale=0.25]{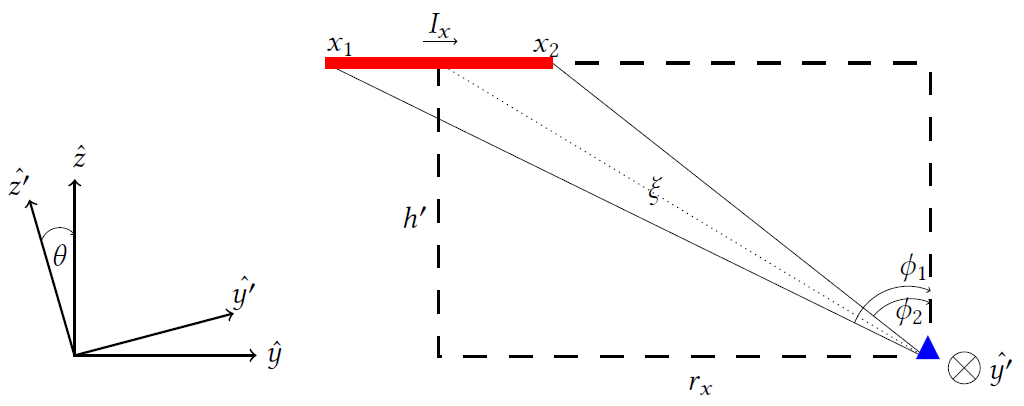}
\caption{In the calculation plane defined by Figure \ref{fig:corr1}, the orthogonal B-field at the ground node can be calculated using the $\phi$ angles that determine the extent of the current element. Projecting the result back to the $\hat{y}$ direction using $\cos(\theta)$ arrives at the resulting horizontal B-field estimate in the correct reference frame.}
\label{fig:corr2}
\end{figure}

For similar representativeness of traditional SECS interpolation, an adapted version of the interpolation scheme can be developed using either short/finite line current elements or infinite line currents to form an equivalent current system (ECS). Physically, we expect a line current basis to be appropriate for the distant magnetospheric current driving seen at mid-latitudes. Using the same underlying approach of SECS, this correction takes the form of a Biot-Savart mapping from each line current element to ground B-field measurement. Assuming a right-handed coordinate system where $z$ is upwards, and making use of the set-up shown in Figure \ref{fig:corr1} and Figure \ref{fig:corr2} we have,

\begin{eqnarray}
B_{x,y}&=&\mp\frac{\mu_0 I_{y,x}}{4\pi h'}\cos(\theta)|\sin(\phi_1)-\sin(\phi_2)|\nonumber\\
&=&\mp\frac{\mu_0I_{y,x}}{4\pi h'}\frac{h}{h'}|\sin(\phi_1)-\sin(\phi_2)|\nonumber\\
&=&\mp\frac{\mu_0I_{y,x}}{4\pi}\frac{h}{r_{x,y}^2+h^2}|\sin(\phi_1)-\sin(\phi_2)|.\label{eq:cor}
\end{eqnarray}

The angles $\phi_1$ and $\phi_2$ are defined as the angles between ground observer and line element end points in a calculation plane ($x',y',z'$), i.e. the plane that includes both the current element and ground observer. Including the orthogonal height $h'$ in the calculation plane results in the magnetic field a distance $\xi$ away from the current element, and defined in the calculation plane reference frame. Since the calculation plane is an angle $\theta$ away from the vertical plane, an additional projection of $\cos(\theta)$ is required to get the magnetic field result back into horizontal components in a standard geographic reference frame. This projection can be simplified to use the calculation plane height $h'$ and standard height $h$. Further simplification allows the entire relation to be defined in terms of the current system height $h$ and the orthogonal projection $r_{x,y}$ (relative to the current element). The final result in \eqref{eq:cor} equates $B_{x,y}$ to orthogonal $I_{y,x}$ current elements, using the $r_{x,y}$ projection orthogonal to the current element and the angles $\phi_1$ and $\phi_2$ which define the extent of the current element. 

In the geometry of the problem, we explicitly have that the $B_{x,y}$ components are mapped to the orthogonal $I_{y,x}$ components (using the appropriate projected displacement components $r_{x,y}$). Note the change of sign for the $-I_y$ current driving $B_x$. Here it should also be stressed that the calculation for the $x$ and $y$ components are done separately, unlike traditional SECS interpolation. The derivation of the Biot-Savart equation takes into account the relation between line element and ground node in the plane formed by the line element and node, with the result then rotated to the typical frame of reference and measurement. As in previous modelling, the distances between current nodes are roughly 100 km and for this case it is assumed that current elements are connected, i.e. the current element lengths either side of each node are half the distance to the adjacent node.  

Instead of assuming discrete line elements in the ECS interpolation scheme, an infinite line current approach can be used in a similar way with projection from a calculation plane. Assuming the line current is infinite in extent compared with the interpolation distances, we have the simplified Biot-Savart form,

\begin{eqnarray}
B_{x,y}=\mp\frac{\mu_0 I_{y,x}}{2 \pi}\frac{h}{r_{x,y}^2+h^2}.\label{eq:cor_inf}
\end{eqnarray}

In the previous implementation of SECS in Southern Africa, with separate calculation of components, the Biot-Savart mapping tends to approximate a $1/r$ drop-off in B-field magnitude when $r\gg h$, i.e. $B_{x,y}=\mu_0I_0/4\pi r_{x,y}$. This implementation is as such able to describe the B-field over a much larger area (1000’s of km) than the scale height of the equivalent current system (100 km). When $r<h$, the divergence-free current system (which can be thought of as a current disc defined for each node) provides horizontal contributions that increasingly cancel each other out, finally cancelling out entirely at $r=0$. The B-field as such is damped by $(1-h/\sqrt{r^2+h^2})$ radially, with a minimum of zero directly under the node (associated with an incident field-aligned current). Physically, this is not expected at mid-latitudes. 

In the line element ECS adaptation presented, when $r\gg h$ the B-field drops off as $1/r^2$, which requires a magnetometer grid with scale length of at most $h$ for reasonable interpolation. Fortunately for mid-latitudes, the dominant driving current systems are magnetospheric currents and $r<h$ for interpolation lengths over Southern Africa (roughly 1000 km). Using an ionospheric equivalent current system defined at a height of 100 km on the other hand results in B-field contributions that drop off significantly. Only scale lengths of roughly 100 km can be correlated, and not the 1,000 km needed over Southern Africa. However, keeping in line with the mid-latitude physical drivers being largely magnetospheric currents, setting the height of the current system at roughly 5,000 km in turn satisfies $r<h$ condition over the region and produces an accurate smooth current system along with accurate B-field interpolation. To stress, an ECS at 5,000 km does not represent a physical current system, but rather an equivalent current system at a height informed by the spacing between magnetometers that correctly captures the far-field contributions of the driving magnetospheric currents over the region of interest.

\section{Results}
\label{sec:res}

To test the different adaptations to SECS, KMH data is kept out-of-sample and used for comparison. SECS and the corrected line current element ECS interpolation scheme use HER, HBK and TSU geomagnetic data to drive interpolation. The dSECS and dECS approaches that use variometers to interpolate $dB/dt$ and update the B-field estimate make use of the SUT, WAT and KMS variometers. To aid comparison, interpolation in all cases uses an equivalent current grid spanning 34.5-18.5°S and 16.5-28.0°E in geographic coordinates. The extent spans the ground measurements and equates to roughly 1,200 km in the east-west direction and 1,700 km in the north-south direction. Using a grid with dimensions of $13\times18$ in these respective directions, a grid spacing of roughly 100 km between current nodes in both directions is obtained. For SECS and dSECS results, an equivalent current height of 100 km is used to match previous implementations in Southern Africa. The line element ECS and dECS instead uses a height of 5,000 km to satisfy the $r<h$ condition, where $r$ is the interpolation scale length.

\begin{figure}
\centering
\includegraphics[scale=0.325]{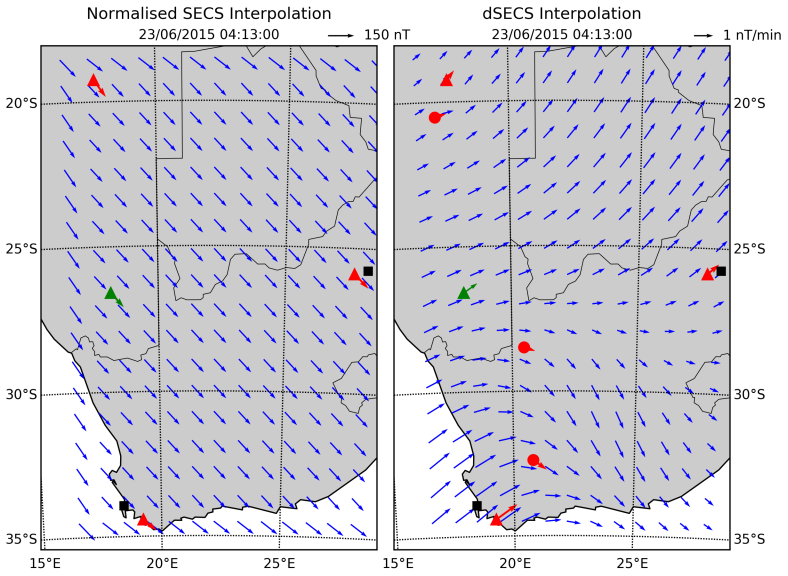}
\caption{Visualisation of SECS and dSECS interpolation over Southern Africa. Red sites drive either the B-field or $dB/dt$ interpolation (blue arrows). KMH (green) is used for validation.}
\label{fig:res1}
\end{figure}

Since both SECS and ECS aim to model equivalent external current systems, the disturbance B-field due to such current systems is what is ultimately interpolated. As such, in pre-processing the data, the background contribution of Earth's intrinsic geomagnetic field is removed. This intrinsic or main field contribution is estimated as the average B-field during preceding geomagnetically quiet time (24 hours). Geomagnetically quiet time in this case is identified by an absolute SYM-H level below 10 nT and ambient solar wind conditions. More sophisticated B-field baseline removal techniques exist, such as using average daily quiet time B-field profile over a month to remove the diurnal solar quiet (\textit{Sq}) variation \citep{Gjerloev2017}. By only removing a constant baseline in our case, the low-level (relative to a geomagnetic storm) \textit{Sq} current system contributions are implicitly included within the disturbance field. A visualisation of the different disturbance fields over Southern Africa using the previous implementation of SECS and dSECS adaptation can be seen in Figure \ref{fig:res1}.

\begin{table}
\centering
\caption{Results of adaptations to B-field interpolation as defined for the KMH validation set during four geomagnetic storms in 2015 ($n=22,224$ data points).}\label{tab:res}
\begin{tabular}{c|c|c}
  Model &   $B_x$ P-score &   $B_y$ P-score \\
\hline
\hline
SECS & 0.83 & 0.80 \\
%\hline
dSECS & 0.91 & 0.88 \\
%\hline
Line element ECS & 0.94 & 0.84 \\
%\hline
Line element dECS & 0.93 & 0.88
\end{tabular}
\end{table}

Table \ref{tab:res} summarises the performance of the various adaptations relative to measured KMH data for the four geomagnetic storms described in Table \ref{tab:storms}. The performance is defined by $P=1-(\textit{RMSE}/\sigma)$ (where \textit{RMSE} is the root mean square error and $\sigma$ is the standard deviation of the measured quantity \citep{Torta2017}). A P-score of 1 reflects zero residuals and indicates perfect modelling. Both dSECS and the corrected line element equivalent current system (ECS) implementations improve on the previous implementation of SECS. $B_x$, which links to the main driver of the magnetospheric east-west currents, sees an improvement of between 10-13\%. The largest improvement, in this case, links to a better representation of the magnetospheric current driving through the line element ECS implementation. Since B-field perturbations at mid-latitudes are dominated by the east-west magnetospheric currents, $B_y$ is more likely affected by induction effects within Earth and finer scale current system perturbations. As such, $B_y$ is less accurately interpolated. That said, $B_y$ shows a similar relative improvement given the use of adapted interpolation (dSECS) with a larger number of B-field measurements through the inclusion of variometers. The combination of the dSECS methodology and line element ECS basis in dECS result reflects the improvements dominated by both current system representation and more localised representation.

\begin{figure}
\centering
\includegraphics[scale=0.51]{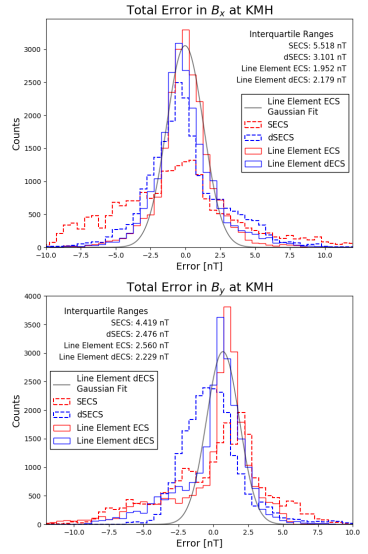}
\caption{Error distributions of adaptations to geomagnetic field interpolation in Southern Africa in terms of the $B_x$ and $B_y$ components at KMH. SECS refers to the previous implementation of SECS by \citet{Bernhardi2008}, and dSECS makes use of the same basis but with the inclusion of variometers and interpolation of $dB/dt$ directly. ECS refers to the alternative line element equivalent current system (ECS), with dECS similarly including variometers and $dB/dt$ interpolation.}
\label{fig:res2}
\end{figure}

A further visualisation of the improvement given adaptions of the interpolation scheme can be seen through the error distributions in Figure \ref{fig:res2}. Using all storms for these total error distributions, it is evident that the proposed adaptations improve B-field interpolation. In both B-field components, the spread of the distribution as defined by the interquartile range is reduced for the dSECS method (44\% improvement). The adaptation error distributions also tend more towards Gaussian (or sharper) error distributions in comparison with the typical SECS method. A more Gaussian profile is suggestive of less systematic error and more random or sampling error. An example of such sampling error would be a known decimal point rounding issue within the KMH dataset used. For the line element ECS implementations, either as is or incorporating the dECS methodology, there is marginal improvement in $B_y$ but large additional improvement in $B_x$. Similar to Table \ref{tab:res}, $B_x$ is linked to the better choice of representative basis function describing dominant east-west magnetospheric currents. Estimation of $B_y$ on the other hand improves with the inclusion of more localized B-field measurements.

The results presented above make use of all available data points during the four geomagnetic storms. When considering the space weather effects of geomagnetic storms, then not all storms are the same. Simply put, larger geomagnetic storms create larger B-field perturbations, which drive larger ground effects. B-field interpolation that aims to estimate equivalent current systems also does better when the driving current systems are stronger. At mid-latitudes, the driving magnetospheric ring currents are directly related to geomagnetic storm magnitude. Considering the period of greatest geomagnetic driving over the available data as such may show improved interpolation accuracy relative to validation that includes quiet time. Using the June 2015 storm data alone, we find higher P-scores of 0.86 and 0.76 relative to using the entire dataset for base SECS interpolation of $B_x$ and $B_y$ respectively. Updating the interpolation to the line element ECS implementation improves the respective P-scores to 0.98 and 0.87. The improved representation of the stronger magnetospheric current driving is very clearly evident in the high  P-score across both implementations. Further, using an infinite line current implementation described in \eqref{eq:cor_inf} results in P-scores of 0.96 and 0.87. It is worth noting that even this simple representation of magnetospheric current driving can result in accurate mid-latitude interpolation during geomagnetic storms.

\begin{figure}
\centering
\includegraphics[scale=0.125]{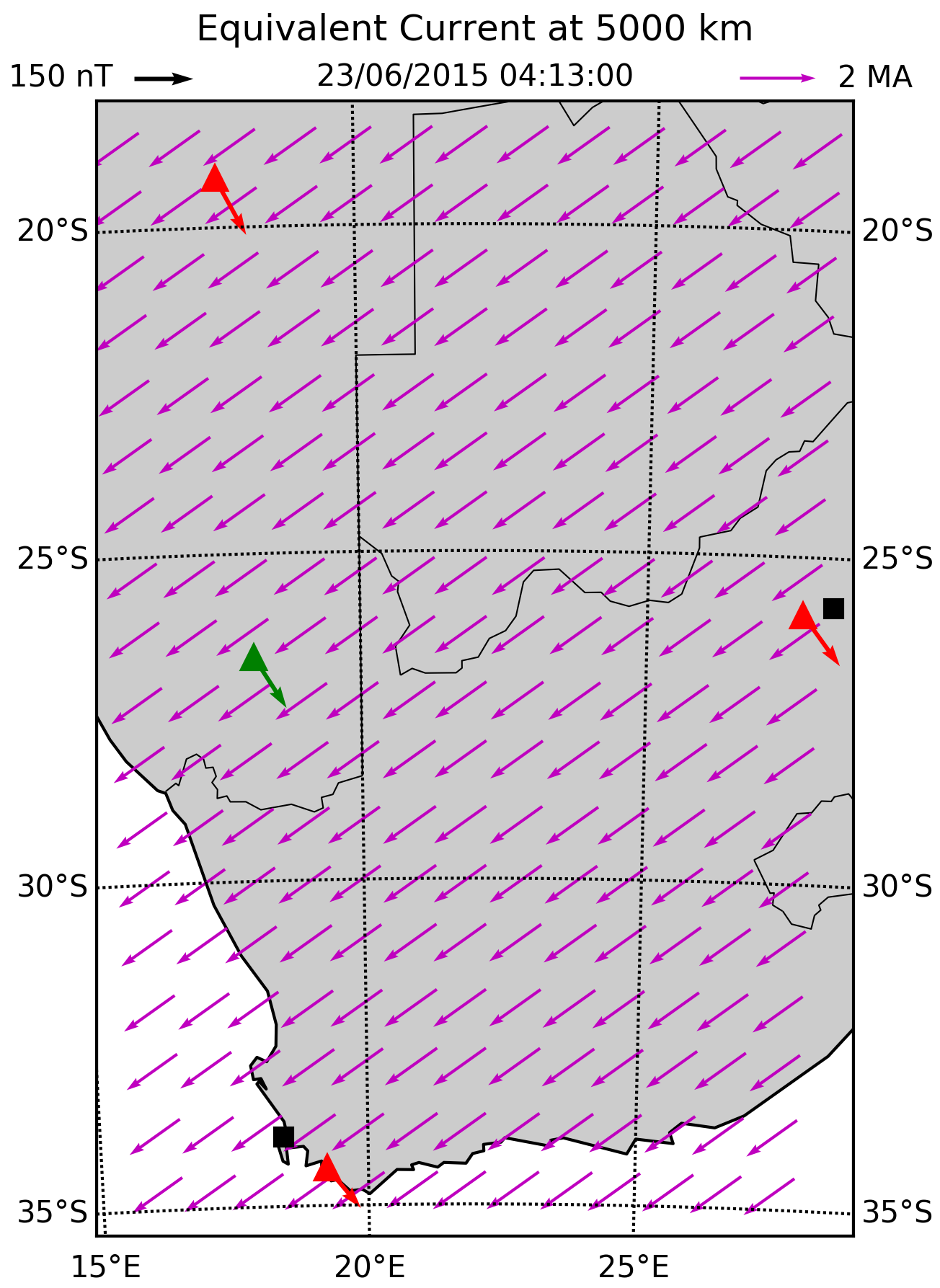}
\caption{An equivalent current system at 5,000 km as determined by line element ECS adaptation of geomagnetic field interpolation. Magenta arrows represent the current system, with the red sites used to drive interpolation and the green arrow indicating the resulting B-field as produced at KMH from the current system. Due to the trade-off between this nearer but more localised equivalent current system relative to magnetospheric currents, care should be taken in interpreting the absolute current strength versus the morphology.}
\label{fig:res3}
\end{figure}

\section{Discussion}
\label{sec:disc}

The adaptations to previous geomagnetic field interpolation proposed in this paper do not nullify the contributions made by previous research using past implementations of SECS in Southern Africa. The validation of the previous implementation of SECS that shows the scheme to be as good as using local variometer measurements supports the technique as a mathematical transfer function. One of the greatest strengths of using a physically consistent equivalent current system for B-field interpolation however lies in its representativeness. In this sense it has not been possible to reconcile a physically relevant mid-latitude equivalent current system to the accuracy of interpolation using the \citet{Bernhardi2008} implementation. 

The need for representativeness has led to the development of the line current element ECS adaptation proposed. Using discrete line current elements, a physically relevant current system surface can be mapped at an appropriate height (as defined by $r<h$, where $r$ is the interpolation scale length). A representation of an equivalent current system as estimated using line current elements is shown in Figure \ref{fig:res3}. In the infinite line current approximation, the various estimates of infinite line currents as defined by current nodes need to be averaged to similarly represent an equivalent current system. 

\begin{figure}
\centering
\includegraphics[scale=0.55]{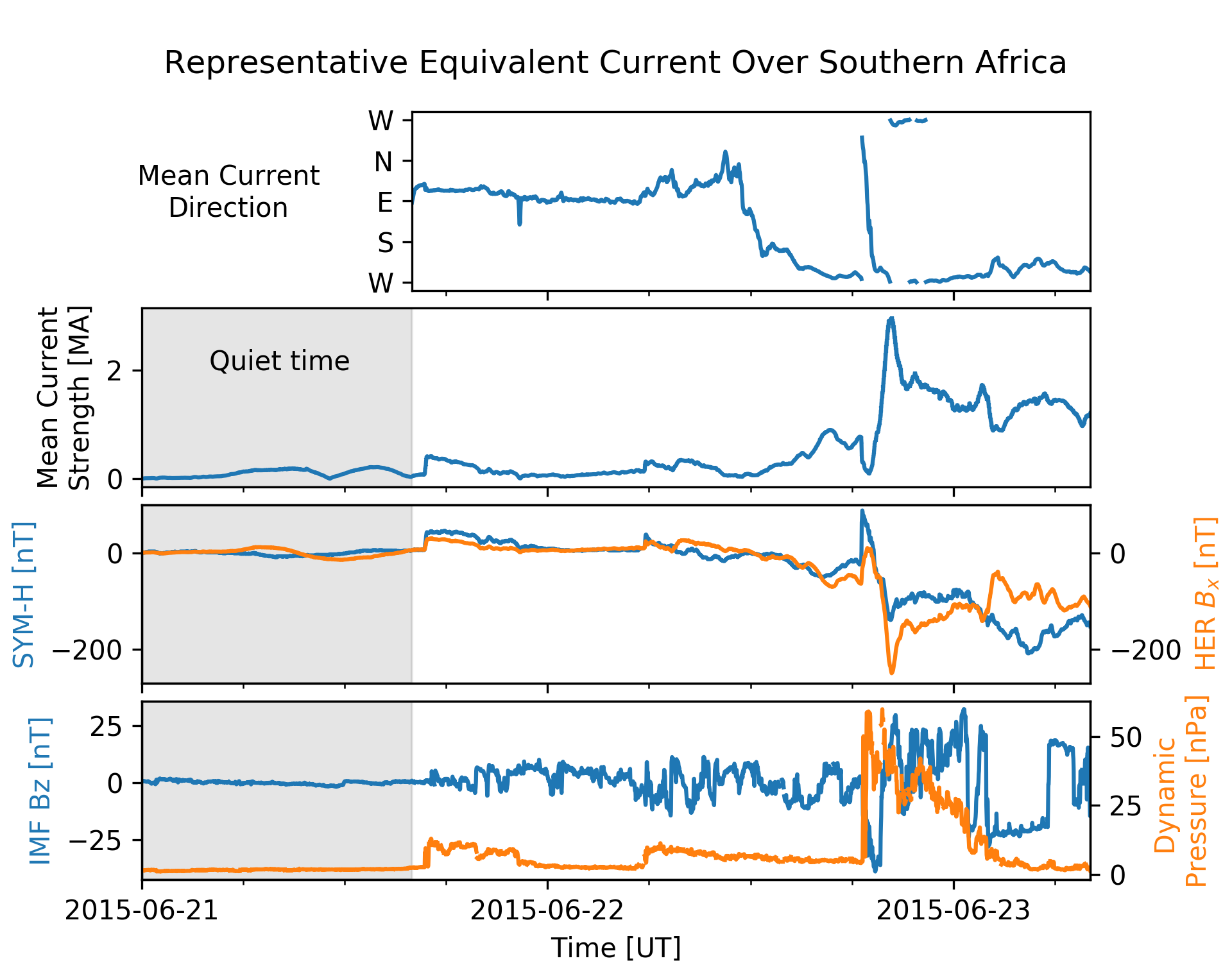}
\caption{Quantifying the equivalent current system in Figure \ref{fig:res3}, the mean current direction and strength is shown in the top two panels. Quiet time fluctuations of the current direction are ignored. In the third panel, the SYM-H index (indicative of the symmetric ring current) is shown along with local $B_x$ measurements at HER. The bottom panel shows in-situ measurements of the $B_z$ component of the interplanetary magnetic field (IMF) and solar wind dynamic pressure at L1.}
\label{fig:res4}
\end{figure}

Extending the ECS representation of Figure \ref{fig:res3}, Figure \ref{fig:res4} shows the mean current element strength and direction, along with solar wind parameters and ground measurements. Evident is the link to the east-west direction of the magnetospheric ring current systems. In this sense we use the term magnetospheric ring current loosely and include the symmetric ring current (a global current system usually associated with the SYM-H index), partial ring current, magnetopause current and tail current. Additional current systems such as the substorm current wedge and field aligned currents add higher order effects \citep{Ganushkina}. The equivalent current system as seen by ground measurements does not distinguish between different current systems and instead is linked to the net effective ground response. During quiet time and ahead of the coronal mass ejection (CME) plasma preceding the geomagnetic storm, the equivalent current magnitude is relatively weak and eastwards, likely related to the symmetric ring current. At onset of the geomagnetic storm, following the arrival of CME plasma and a southward (-$B_z$) interplanetary magnetic field (IMF) that initiates reconnection and particle injection, the westward magnetospheric current increases dramatically as expected. The strengthening of the net westward equivalent current corresponds to contributions first from tail current in the early main phase and then from the partial ring current, before giving rise to a strong symmetric westward ring current in the recovery phase \citep{Ganushkina}.

Given the combination of sparse magnetometers (with a spacing of around 1,000 km in Southern Africa) and driving from distant magnetospheric currents, an equivalent current system height on the order of 5,000 km ($h$ being greater than the interpolation distances $r$) accurately interpolates the B-field over Southern Africa. As the height parameter is dependent on magnetometer spacing, for a denser magnetometer grid the current system height can be reduced. At mid-latitudes the reduced height and better resolution is unlikely to provide much benefit as the driving current systems remain dominated by magnetospheric currents that sit between 10,000 and 60,000 km \citep{Daglis1999,Ganushkina}. In testing the line element ECS interpolation approach, reflecting an actual magnetospheric height (i.e. $h\gg$ 5,000 km) provides only marginal improvement compared to $h=$ 5,000 km. In contrast, there were significant improvements in performance with increasing height from $h=$ 100 km, at which height there was no correlation between magnetometer sites and interpolation broke down entirely. At higher latitudes, where ionospheric auroral currents dominate, a denser magnetometer grid and lower height would be needed to maintain performance of the line elements implementation. Here the use of representative divergence-free current nodes instead of discrete line current elements greatly improves the efficiency of magnetometer measurements used and more accurately represents the current system dynamics at play. The traditional approach to SECS is as such best suited for high-latitudes, and the line element adaptation allows the equivalent current system approach of SECS interpolation to be used at mid-latitudes, such as in Southern Africa. Further work is required to validate the line element approach at high- and low-latitudes and determine where it possibly breaks down.

The proposed adaptation of including variometers in geomagnetic field interpolation has direct operational application. The dSECS methodology is independent of the basis functions used and focuses on interpolating $dB/dt$ directly, then merging the result with an initial B-field interpolation. The rate of change $dB/dt$ is extensively used in magnetotelluric (MT) studies, with $dB/dt$ related to the B-field by a factor $i\omega$ in the frequency domain. In modelling geomagnetically induced currents (GICs), frequency components need to be weighted correctly to reproduce GIC effective phenomena, like low-frequency geomagnetic pulsations \citep{Heyns2020}. To do this, a low-pass filter related to the conductivity structure of the Earth needs to be applied to $dB/dt$ (which otherwise would be biased to the sampling rate of the underlying B-field measurements). With these aspects taken into account, $dB/dt$ interpolation can in effect be used as is. A caveat in $dB/dt$ interpolation is that when line element equivalent currents are used at an appropriate height, they are less sensitive to local variation in the B-field, i.e. contributions are dominated by magnetospheric currents which produces a smoother $dB/dt$ interpolation map but still an accurate B-field interpolation map. 

With the increased need for real-time monitoring and modelling of GICs by utilities, dSECS and dECS provide an attractive approach to B-field estimation. The roll-out of low-cost instrumentation by a utility is much more feasible than installation and maintenance of observatory-grade magnetometers and additionally allows data access without the need for third party involvement. The low-cost of variometers has further application in research and citizen science \citep{Beggan2016}. Although these instruments lack accuracy individually, leveraging a collection of them to estimate an equivalent current system can greatly improve estimation of the B-field at any single site, even at those sites where a variometer is situated (which in turn has application in validation of measurements). As more variometers are included in interpolation, the more accurate the resulting B-field estimations become. A secondary output of the dSECS methodology presented is the use an improved estimation of variation in a parameter to efficiently constrain a baseline estimate of the same parameter. Many fields face similar challenges where the variation in a parameter over time is more certain than its baseline estimation, making the approach presented more generally applicable.

\section{Conclusion}
\label{sec:con}

Due to the costs associated with maintaining observatory-grade magnetometers, it is not feasible to generate dense maps of geomagnetic field measurements. Interpolation of the geomagnetic field provides an attractive alternative that is widely used in research and operational contexts. Aside from the improved accuracy needed for operational modelling of geomagnetically induced currents, using a physically inspired interpolation scheme allows the equivalent currents to be interpreted. This paper presents two novel adaptations of previous geomagnetic field interpolation in Southern Africa. First, more widely available low-cost variometers are included to provide more accurate interpolation. A direct implication is that utilities and geomagnetic observatories can supplement their existing magnetometers for improved geomagnetic field estimation at a fraction of the cost. Secondly, an equivalent current system interpolation scheme using line current elements was developed for use at mid-latitudes. This approach is more representative of the driving current system at mid-latitudes compared with the more generally used divergence-free spherical elementary current systems (SECS). The implementation further allows simple analysis of the equivalent currents, not previously possible with an incorrect implementation of SECS in Southern Africa.

\section*{Acknowledgments}
The results presented in this paper rely on the data collected at Hermanus, Hartebeesthoek, Tsumeb and Keetmanshoop geomagnetic observatories. We thank SANSA for supporting their operation and INTERMAGNET (\url{https://www.intermagnet.org}) for promoting high standards of magnetic observatory practice. Further data from pulsation magnetometers and magnetotelluric sites operated by SANSA were used and are available through SANDIMS (\url{https://sandims.sansa.org.za/}). GFZ K-index data can be accessed through \url{https://doi.org/10.5880/Kp.0001}. The SYM-H index and interplanetary magnetic field measurements can be accessed through NASA/GSFC Space Physics Data Facility’s OMNIWeb (or CDAWeb or ftp) service (\url{https://omniweb.gsfc.nasa.gov/}). This work was funded in part by a grant from the Open Philanthropy Project Fund, University of Cape Town and the South African National Space Agency.

%% Bibliography
%% Author year style
\bibliographystyle{model5-names}
\biboptions{authoryear}
%\bibliography{../library}

\end{document}